\documentclass[prc,twocolumn,nofootinbib,superscriptaddress,showpacs,amssymb,amsmath,amsfonts,aps]{revtex4}
\usepackage{graphicx}
\usepackage{dcolumn}
\begin{document}
\title{Electroproduction of $\phi(1020)$ mesons at $1.4\leq Q^2\leq$ GeV$^2$ measured with the CLAS spectrometer}

\newcommand*{\VT}{Virginia Polytechnic Institute and State University, Blacksburg, Virginia   24061-0435}
\affiliation{\VT}
\newcommand*{\CUA}{Catholic University of America, Washington, D.C. 20064}
\affiliation{\CUA}
\newcommand*{\JLAB}{Thomas Jefferson National Accelerator Facility, Newport News, Virginia 23606}
\affiliation{\JLAB}
\newcommand*{\SACLAY}{CEA-Saclay, Service de Physique Nucl\'eaire, 91191 Gif-sur-Yvette, France}
\affiliation{\SACLAY}
\newcommand*{\ORSAY}{Institut de Physique Nucleaire ORSAY, Orsay, France}
\affiliation{\ORSAY}
\newcommand*{\ANL}{Argonne National Laboratory}
\affiliation{\ANL}
\newcommand*{\ASU}{Arizona State University, Tempe, Arizona 85287-1504}
\affiliation{\ASU}
\newcommand*{\UCLA}{University of California at Los Angeles, Los Angeles, California  90095-1547}
\affiliation{\UCLA}
\newcommand*{\CSU}{California State University, Dominguez Hills, Carson, CA 90747}
\affiliation{\CSU}
\newcommand*{\CMU}{Carnegie Mellon University, Pittsburgh, Pennsylvania 15213}
\affiliation{\CMU}
\newcommand*{\CNU}{Christopher Newport University, Newport News, Virginia 23606}
\affiliation{\CNU}
\newcommand*{\UCONN}{University of Connecticut, Storrs, Connecticut 06269}
\affiliation{\UCONN}
\newcommand*{\ECOSSEE}{Edinburgh University, Edinburgh EH9 3JZ, United Kingdom}
\affiliation{\ECOSSEE}
\newcommand*{\EMMY}{Emmy-Noether Foundation, Germany}
\affiliation{\EMMY}
\newcommand*{\FU}{Fairfield University, Fairfield CT 06824}
\affiliation{\FU}
\newcommand*{\FIU}{Florida International University, Miami, Florida 33199}
\affiliation{\FIU}
\newcommand*{\FSU}{Florida State University, Tallahassee, Florida 32306}
\affiliation{\FSU}
\newcommand*{\GWU}{The George Washington University, Washington, DC 20052}
\affiliation{\GWU}
\newcommand*{\ECOSSEG}{University of Glasgow, Glasgow G12 8QQ, United Kingdom}
\affiliation{\ECOSSEG}
\newcommand*{\ISU}{Idaho State University, Pocatello, Idaho 83209}
\affiliation{\ISU}
\newcommand*{\INFNFR}{INFN, Laboratori Nazionali di Frascati, 00044 Frascati, Italy}
\affiliation{\INFNFR}
\newcommand*{\INFNGE}{INFN, Sezione di Genova, 16146 Genova, Italy}
\affiliation{\INFNGE}
\newcommand*{\BONN}{Institute f\"{u}r Strahlen und Kernphysik, Universit\"{a}t Bonn, Germany}
\affiliation{\BONN}
\newcommand*{\ITEP}{Institute of Theoretical and Experimental Physics, Moscow, 117259, Russia}
\affiliation{\ITEP}
\newcommand*{\JMU}{James Madison University, Harrisonburg, Virginia 22807}
\affiliation{\JMU}
\newcommand*{\KYUNGPOOK}{Kyungpook National University, Daegu 702-701, Republic of Korea}
\affiliation{\KYUNGPOOK}
\newcommand*{\MIT}{Massachusetts Institute of Technology, Cambridge, Massachusetts  02139-4307}
\affiliation{\MIT}
\newcommand*{\UMASS}{University of Massachusetts, Amherst, Massachusetts  01003}
\affiliation{\UMASS}
\newcommand*{\MOSCOW}{Moscow State University, General Nuclear Physics Institute, 119899 Moscow, Russia}
\affiliation{\MOSCOW}
\newcommand*{\UNH}{University of New Hampshire, Durham, New Hampshire 03824-3568}
\affiliation{\UNH}
\newcommand*{\NSU}{Norfolk State University, Norfolk, Virginia 23504}
\affiliation{\NSU}
\newcommand*{\OHIOU}{Ohio University, Athens, Ohio  45701}
\affiliation{\OHIOU}
\newcommand*{\ODU}{Old Dominion University, Norfolk, Virginia 23529}
\affiliation{\ODU}
\newcommand*{\PITT}{University of Pittsburgh, Pittsburgh, Pennsylvania 15260}
\affiliation{\PITT}
\newcommand*{\RPI}{Rensselaer Polytechnic Institute, Troy, New York 12180-3590}
\affiliation{\RPI}
\newcommand*{\RICE}{Rice University, Houston, Texas 77005-1892}
\affiliation{\RICE}
\newcommand*{\URICH}{University of Richmond, Richmond, Virginia 23173}
\affiliation{\URICH}
\newcommand*{\SANTAMARIA}{Universidad T\'ecnica Federico Santa Mar\'{\i}a, Casilla 110-V, Valpara\'\i so, Chile}
\affiliation{\SANTAMARIA}
\newcommand*{\SCAROLINA}{University of South Carolina, Columbia, South Carolina 29208}
\affiliation{\SCAROLINA}
\newcommand*{\UNIONC}{Union College, Schenectady, NY 12308}
\affiliation{\UNIONC}
\newcommand*{\VIRGINIA}{University of Virginia, Charlottesville, Virginia 22901}
\affiliation{\VIRGINIA}
\newcommand*{\WM}{College of William and Mary, Williamsburg, Virginia 23187-8795}
\affiliation{\WM}
\newcommand*{\YEREVAN}{Yerevan Physics Institute, 375036 Yerevan, Armenia}
\affiliation{\YEREVAN}
\newcommand*{\NOWOHIOU}{Ohio University, Athens, Ohio  45701}
\newcommand*{\NOWJLAB}{Thomas Jefferson National Accelerator Facility, Newport News, Virginia 23606}
\newcommand*{\NOWUNH}{University of New Hampshire, Durham, New Hampshire 03824-3568}
\newcommand*{\NOWCNU}{Christopher Newport University, Newport News, Virginia 23606}
\newcommand*{\NOWGWU}{The George Washington University, Washington, DC 20052}
\newcommand*{\NOWUMASS}{University of Massachusetts, Amherst, Massachusetts  01003}
\newcommand*{\NOWMIT}{Massachusetts Institute of Technology, Cambridge, Massachusetts  02139-4307}
\newcommand*{\NOWURICH}{University of Richmond, Richmond, Virginia 23173}
\newcommand*{\NOWECOSSEE}{Edinburgh University, Edinburgh EH9 3JZ, United Kingdom}
\newcommand*{\NOWGEISSEN}{Physikalisches Institut der Universitaet Giessen, 35392 Giessen, Germany}
\newcommand*{\NOWDECEASED}{Deceased}

\author {J.P.~Santoro}
\affiliation{\VT}
\affiliation{\CUA}
\author {E.S.~Smith} 
\affiliation{\JLAB}
\author {M.~Gar\c con} 
\affiliation{\SACLAY}
\author {M.~Guidal} 
\affiliation{\ORSAY}
\author {J.M.~Laget} 
\affiliation{\JLAB}
\author {C.~Weiss} 
\affiliation{\JLAB}

\author {G.~Adams} 
\affiliation{\RPI}
\author {M.J.~Amaryan} 
\affiliation{\ODU}
\author {M.~Anghinolfi} 
\affiliation{\INFNGE}
\author {G.~Asryan} 
\affiliation{\YEREVAN}
% ...

\collaboration{The CLAS Collaboration}

\begin{abstract}

Electroproduction of exclusive $\phi$ vector mesons has been studied
with the CLAS detector in the kinematical range $1.4\leq Q^2\leq 3.8$
GeV$^{2}$, $0.0\leq t^{\prime}\leq 3.6$ GeV$^{2}$, and $2.0\leq W\leq
3.0$ GeV.  The scaling exponent for the total cross section as
$1/(Q^2+M_{\phi}^2)^n$ was determined to be $n=2.49\pm 0.33$. The
slope of the four-momentum transfer $t'$ distribution is
$b_{\phi}=0.98 \pm 0.17$ GeV$^{-2}$.  Under the 
assumption of s-channel helicity conservation (SCHC),
we determine the ratio of longitudinal to transverse cross
sections to be $R=0.86 \pm 0.24$. A 2-gluon exchange model is
able to reproduce the main features of the data.

\end{abstract} 
\maketitle

\section{Introduction}

Exclusive electroproduction of vector mesons is an essential tool 
for exploring the structure of the nucleon and the exchange mechanisms 
governing high--energy scattering. For low photon virtualities relative to
the vector meson mass, 
$Q^2 \lesssim m_V^2$, or in the case of photoproduction, $Q^2 = 0$, 
these processes are well described by $t$--channel exchange of Regge poles
(Pomeron, Reggeon) --- extended objects whose properties can be related 
to the observed hadron spectrum \cite{JMLaget2}. At high virtualities, 
$Q^2 \gg m_V^2$, a QCD factorization theorem \cite{Collins} states that 
vector meson production from longitudinally polarized photons proceeds 
by exchange of a small--size system of quarks or gluons, whose coupling 
to the nucleon is described by the generalized parton distributions (GPDs). 
By studying the dependence of exclusive electroproduction on $Q^2$, one can 
thus ``resolve'' the Pomeron and Reggeon into their quark and gluon 
constituents. Additional information comes from the comparison of 
the $\rho^0, \omega$ and $\phi$ channels, which couple differently to
quarks and gluons. The self-analyzing decays of the spin--1 mesons 
allow one to study also the helicity structure of the $\gamma^\ast N$ 
interaction and, assuming helicity conservation, to separate longitudinal and transverse
photon polarizations.

This article presents data for exclusive $\phi$ vector meson
electroproduction off the proton above the resonance region,
taken with a 5.754 GeV electron beam of the CEBAF accelerator and 
the CLAS detector at Jefferson Lab \cite{CLAS_nim_paper}.
The measurement was performed as part of a series of experiments
aimed at studying vector meson production in the valence quark region 
at the highest available photon virtualities. The analysis of $\omega$ 
production has been completed \cite{ludyvine_paper}, and the analysis 
of $\rho$ production is in progress \cite{Guidal}. The analysis of
$\phi$-meson production reported here is based in part on the work  
of Ref.\cite{myThesis}.  

The $\phi$-meson is unique in that its quark composition is mostly $\bar ss$ 
containing little, if any, $u$ and $d$ flavors which populate the valence quarks in 
the nucleon. Thus, $\phi$ production primarily probes the gluon 
degrees of freedom in the target. High--energy photoproduction of 
$\phi$ proceeds mainly by Pomeron exchange. At large $Q^2$, calculations
based on current GPD models show that the $\phi$ production cross 
section is dominated by the gluon GPD, with only small contributions 
arising from intrinsic strange quarks in the nucleon \cite{Diehl:2005gn,Goloskokov:2006hr}. 
At intermediate $Q^2$, a description of $\phi$ production based on 
effective two--gluon exchange has been proposed \cite{JMLaget1}, 
which effectively interpolates between the ``soft'' and ``hard'' regimes. 
Thus, $\phi$ production provides us with a clean method of probing the 
gluon field in the nucleon, even at JLab energies. 

A natural framework for discussing exclusive vector meson production
is the space--time picture in the target rest frame (\textit{i.e.}, 
the laboratory frame) \cite{bauer}.
At high energies, the interaction 
of the virtual photon with the target proton proceeds by way of 
fluctuation of the photon into virtual hadronic (or quark--antiquark)
configurations that subsequently scatter diffractively off the target.
This process occurs over a characteristic time given by the
lifetime of the fluctuation as dictated by the uncertainty
principle and is given by
\begin{eqnarray}
\Delta\tau & = & {2\nu\over(Q^{2}+M_{\text{fluct}}^{2})} ,
\label{DelTau}
\end{eqnarray}
where $\nu$ is the photon laboratory energy and $M_{\text{fluct}}$  is the mass 
of the virtual hadronic state. This interval also determines 
the coherence length  in the longitudinal direction,
$l_{\text{coh}} = c\Delta\tau$. 
In photoproduction or electroproduction at $Q^2 \lesssim m_V^2$, 
this picture is the basis for the successful vector dominance model (VDM), 
where the dominant hadronic fluctuations are assumed to be the observed 
ground--state vector mesons ($\rho^0, \omega, \phi$). Their interaction
with the target can be described by Pomeron exchange. As $Q^2$ increases, 
higher--mass states become important. Eventually, at $Q^2 \gg m_V^2$, 
the fluctuations of the photon can appropriately be described as 
quark--antiquark pairs (``dipoles'') with transverse momenta 
$k_\perp^2 \sim Q^2$, or transverse size $r_\perp \sim 1/Q \ll 1/m_V$. 
Their interaction with the nucleon is described by the gluon GPD, 
which can be interpreted as the ``color dipole moment'' of the target.

In the context of the space--time picture, measuring the $Q^2$--dependence
of exclusive electroproduction up to $Q^2 \sim \text{few GeV}^2$ allows one 
to vary the transverse size of the projectile from ``hadronic size'' 
($r_\perp \sim 1/m_V$) to ``small size'' ($r_\perp \sim 1/Q$), thus 
resolving the structure of the target at very different distance scales.
At HERA energies, where $l_{\text{coh}} \gg 1 \, \text{fm}$ even for
$Q^2 \sim \text{few GeV}^2$, one can neglect the variation of the 
coherence length with $Q^2$ and associate the $Q^2$--dependence 
entirely with a change of the transverse size of the projectile.
The predictions derived in this approximation are nicely confirmed
by the data, \textit{e.g.} the decrease of the $t$--slope with $Q^2$,
and the increase with $Q^2$ of the exponent governing the energy 
dependence (for a review see Ref.~\cite{Frankfurt:2005mc}). 
At JLab energies, where the coherence length in 
electroproduction is $l_{\text{coh}} \lesssim 1 \, \text{fm}$,
we must also take into account its variation with $Q^2$, \textit{i.e.},
the ``shrinkage'' of the longitudinal size of the virtual photon
with increasing $Q^2$. Another effect modifying the space--time
interpretation is the non-negligible longitudinal momentum transfer 
to the target, which increases with $Q^2$.
Nevertheless, the space--time picture remains a very useful
framework for discussing vector meson electroproduction even at
JLab energies.

In the present $\phi$-meson production experiment, the $t$--dependence 
of the differential cross section was measured over a wide range,
from the kinematic minimum at $t \sim t_{0}$ (small CM scattering angle) to 
$t \sim s/2$ (large angle). In exclusive electroproduction
$t$ is related to the transverse momentum transfer to the target,
$\Delta_\perp^2$, and thus
determines the effective impact parameters in the cross section, 
$b_\perp \sim 1/\Delta_\perp$ \footnote{The
transverse momentum transfer to the target is given by
$\Delta_\perp^2 = (1 - \xi^2) (t - t_{0})$,
where $\xi$ is
the fractional longitudinal momentum transfer to the target, which in turn is
related to the Bjorken variable in the kinematics of
deep--inelastic scattering, $\xi = x_B / (2 - x_B)$.}.
Exclusive meson production at large $-(t - t_{0})$ probes
configurations of small transverse size in the target. 
The possibility to vary both $Q^2$ and
$t$ in electroproduction allows one to control both the size
of the projectile and the size of the target configurations
contributing to the process, and to study their interplay \cite{JMLaget3}.

Quantitative predictions for the production of vector mesons in our 
kinematic regime have been made by Laget and collaborators based on the 
interactions between constituent partons (JML model). The high-$t$ 
behavior of the photoproduction cross section of $\phi$-mesons \cite{anciant-2000-85} has 
been reproduced using dressed gluon propagators  and correlated quark 
wave functions in the proton \cite{CANOJMLaget2}. Quark exchange processes, which
contribute also to the photoproduction of $\rho$ and $\omega$ mesons, have 
been modeled in terms of saturating Regge trajectories. The 
model uses electromagnetic form factors in the Regge amplitude \cite{CANOJMLaget,JMLaget3} 
to describe electroproduction data. However, the $Q^2$ dependence 
of the 2-gluon amplitude is an intrinsic part of its construction, and 
no additional electromagnetic form factors are needed. Therefore, the 
predicted $\phi$-meson electroproduction cross section is parameter free 
and constitutes a strong test of the partonic description that underlies 
the model. The full form for the amplitudes are given in 
Refs.\,\cite{JMLaget2,JMLaget1,CANOJMLaget2}. Thus far, comparisons 
of the JML model for electroproduction have been
made with $\omega$ \cite{omegaTHESIS,ludyvine_paper}, and $\rho$ \cite{Guidal,cynthia_paper}
electroproduction data from JLab, and $\rho$ electroproduction data
from HERMES \cite{HERMES}.

One of the leading motivations for the present work is the sparse
amount of existing $\phi$ electroproduction data.  The body of
$\phi$-meson electroprodution data at similar kinematics consists of
early data from Cornell \cite{Dixon,Dixon2,Cassel}, and some data from CLAS
at lower energy \cite{costypaper}. Recent data on $\phi$
electroproduction comes from HERMES \cite{Borissov,HERMES} and
HERA \cite{Adloff2,Adloff1,Breitweg,ZEUS_HERA_2005} at much higher center-of-mass energy ($W$). 
A summary of the world data indicating their kinematic range is given in Table\,
\ref{datasummary}. The data from this experiment are complementary to
measurements at collider energies which cover a higher $W$ and higher
$Q^2$ range where diffraction mechanisms are probed.


\begin{table}[h]
\caption{Summary of $\phi$ electroproduction data and kinematic range.}
\begin{center} \begin{tabular}{|l|c|c|} \hline
  Experiment                        & $Q^2$ (GeV$^2$)                & $W$ (GeV)                    \\ \hline
  Cornell Dixon \cite{Dixon,Dixon2} & 0.23 - 0.97  & 2.9             \\
  Cornell Cassel \cite{Cassel}      & 0.80 - 4.00 & 2.0 - 3.7        \\
  HERMES \cite{Borissov,HERMES}     & 0.70 - 5.00 & 4.0 - 6.0        \\
  CLAS\cite{costypaper}             & 0.70 - 2.20 & 2.0 - 2.6        \\ \hline
  H1 \cite{Adloff2}                 & $>$ 7.0 &  $\sim$ 75.0         \\
  H1 \cite{Adloff1}                 & 1.00 - 15.0 & 40.0 - 130.0     \\
  H1 \cite{Breitweg}                & 3.00 - 20.0 & 4.0 - 120.0      \\
  ZEUS \cite{ZEUS_HERA_2005}        & 7.00 - 25.0 & 42.0 - 134.0     \\ 
  ZEUS \cite{ZEUS_HERA_2005}        & 2.00 - 70.0 & 35-145       \\ \hline
\end{tabular}
\label{datasummary}
\end{center}
\end{table}

We have measured $\phi$-meson
electroproduction at the highest possible $Q^2$ accessible at CEBAF energies in the
valence quark regime. 
The data set covers the kinematical regime
$1.4\leq Q^2\leq3.8$ GeV$^2$, $0.0\leq t^{\prime}\leq 3.6$ GeV$^2$, and
$2.0\leq W\leq 3.0$ GeV. We will present cross sections as a function of
the momentum transfer $-t$, the azimuthal angle $\Phi$ between the
electron and hadron scattering planes, as well as the angular decay
distributions in the rest frame of the $\phi$-meson. Although
limitations of the statistical sample will preclude determining
correlations between different kinematic variables, the distributions will provide
insights into the distance scale of the interaction and explore kinematics
that begin to probe partonic degrees of freedom.
 
\section{Kinematics and notation}
\begin{center}
\begin{figure}[t]
\includegraphics[width=8cm,clip=true,bb=6 109 460 460]{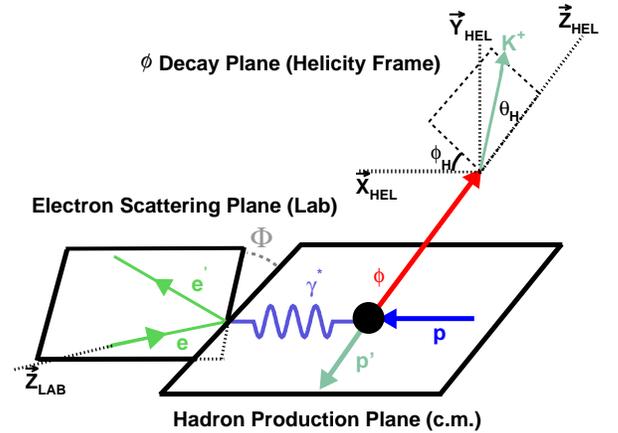}
\caption{\label{electropro} (Color online) Graphical representation of $\phi$-meson 
electroproduction. Shown from left to right then above are 
the electron scattering plane, the hadron production plane and helicity rest 
frame of the $\phi$ respectively. }
\end{figure}
\end{center}

The kinematic variables in exclusive
$\phi$ production (see Fig.\,\ref{electropro}) described by 
\begin{eqnarray}
e(k)\,p(P) \rightarrow e(k')\, \phi(\upsilon)\, p(P'),
\end{eqnarray}
are $k$, $k'$, $P$, $P'$ and $\upsilon$ which are, respectively, the four-momenta
of the incident electron, scattered electron, target proton, scattered
proton and the $\phi$-meson:
\begin{itemize}
\item $Q^2$ = $-q^2$ = $-(k-k')^2$, the negative four-momentum 
squared of the virtual photon;
\item $W^2$ = $(q+P)^2$, the squared invariant mass of the photon-proton
system;
\item $x_B$ = $Q^2/(2P\cdot q)$, the Bjorken scaling variable;
\item $\nu$ = $P\cdot q /M_p$, energy of the virtual photon;  
\item $t$ = $(P-P')^2$, the squared four-momentum transfer at the proton
vertex, is given by
\begin{eqnarray}
t & = & t_{0} - 4p^{\gamma^*}_{cm} p^{\phi}_{cm} \sin^2(\theta_{cm}/2)\; {\rm , where} \nonumber \\
t_{0} & = & ({{E}^{\gamma^{*}}_{cm}}-{{E}^{\phi}_{cm}})^2
-(p^{\gamma^*}_{cm}- p^{\phi}_{cm})^2 \nonumber
\end{eqnarray} 
and the above formulas are calculated 
using the energy and momenta of
the virtual photon and $\phi$ in the $\gamma^* p$ center-of-mass;

\item $t'$ = $|t - t_{0}|$, momentum transfer relative to the kinematic
limit $-t_0$, which increases with Q$^2$ and decreases with increasing W;
\item The coordinate system in the $\gamma^* p$ center-of-mass is defined
with the z-axis along the direction of the virtual photon, and the y-axis
normal to the hadronic production plane along $\vec{p}\;^{\gamma^*}_{cm} \times \vec{p}\;^{\phi}_{cm} $;
\item $\Phi$, the angle between the hadron production ($\gamma^*\phi p$) plane
and the electron scattering ($ee'\gamma^*$) plane
following the convention in Ref. \cite{wolfShilling}\footnote{The azimuthal
angle $\Phi$ used here is $-\phi$ from the ``Trento convention'' \cite{Diehl:2007jy}.};
\item $\cos{\theta_{H}}$ and  $\phi_H$, decay angles 
of the $K^+$ in the helicity frame  \cite{wolfShilling}, which is defined 
in the rest frame of the $\phi$-meson with the z-axis along the
direction of the $\phi$-meson in the $\gamma^* p$ center-of-mass system;
\item $\psi$ = $\phi_H - \Phi$, azimuthal angle that simplifies
the angular decay distributions when s-channel helicity is conserved (SCHC).
\end{itemize}

The electroproduction reaction integrated over the decay angles of 
the $\phi$-meson can be described by the following set of four independent variables:
$Q^2$, $-t$, $\Phi$ and $W$. For the analysis of the decay distribution, 
the additional variables $\cos{\theta_H}$ and $\psi$ are required.
In total there are six independent variables in the approximation of negligible
$\phi$ width.

\section{Experiment}
The experiment was conducted with the CEBAF Large Acceptance
Spectrometer (CLAS) \cite{CLAS_nim_paper} located in Hall B of the Thomas
Jefferson National Accelerator Facility.  The CLAS spectrometer is
built around six independent superconducting coils that generate a
toroidal magnetic field azimuthally around the beam direction.  The
azimuthal coverage is limited by the magnetic coils and is
approximately 90\% at large angles and narrows to 50\% at forward
angles. Each sector is equipped with three regions of multi-wire
drift chambers and time-of-flight counters that cover the angular
range from 8$^{\circ}$ to 143$^{\circ}$. Charged-particle trajectories are
tracked through the field with the drift chambers, and the
scintillators provide a precise determination of the particle flight
time.  In the forward region (8$^{\circ}$ to 45$^{\circ}$), each sector is
furthermore equipped with gas-filled threshold Cerenkov counters (CC) and
electromagnetic calorimeters (EC).  The Cerenkov counters are used to
discriminate electrons from pions, and the calorimeters are used to
measure the energy of electrons and photons.

The data were collected between October 2001 and January 2002 with a
5.754 GeV electron beam incident on a 5 cm-long liquid hydrogen
target. The typical beam current was 7 nA.  The CLAS torus magnet
was set to 3375 A with a polarity that caused negatively charged
particles to bend in towards the beamline. The inclusive electron
trigger fired when signals in the forward electromagnetic calorimeter
exceeded a predefined threshold in coincidence with a hit in the
Cerenkov counters.  The kinematical domain of the selected sample
corresponds approximately to $Q^2$ from 1.5 to 5.5 GeV$^2$ and $W$
between 2 and 3 GeV.  The typical experimental dead time was about 8\% 
with a trigger rate of about 1.5 kHz.

\section{Event Reconstruction}
The $\phi$-mesons were detected using the charged-particle decay mode into
$K^{+}$ and a $K^{-}$.  Events corresponding to
$ep\rightarrow epK^{+}(K^{-})$ were classified initially by requiring at
least one negative track and two positive tracks. Normally the $K^-$
remained undetected due to the limited acceptance for negative particles
at this high magnetic field setting. After calibration of
the spectrometer, the momentum of each particle was determined 
with a fractional resolution of about a percent using
the track segments in the drift chambers. The momentum resolution is 
sufficient to identify the missing particle as a $K^-$.

\begin{figure}[t]
 \includegraphics[height=6cm,clip=true,bb=5 5 560 430]{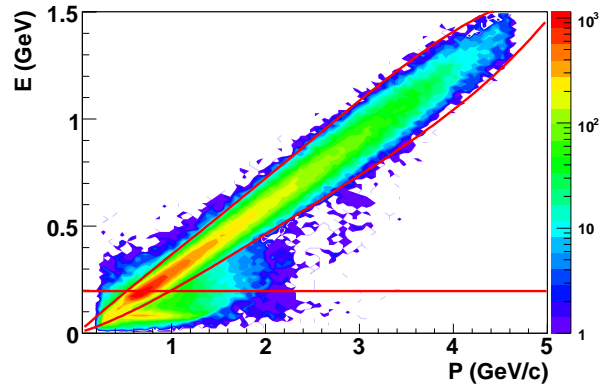}
 \caption{\label{eoverpvsp1}Energy deposited by the electron candidates
in the electromagnetic calorimeter versus momentum.
The lines show the selection cuts for good electrons as described in the text.
 }
 \end{figure}

The identification of good electrons is the crucial first step and is
accomplished through energy and momentum cuts \cite{myThesis}. After
selection of tracks within the fiducial volume of the detector, the
momentum of the electron candidate track in each event was required to correspond to the
energy deposition in the electromagnetic calorimeter and the visible
energy be greater than 0.2 GeV (Fig.\,\ref{eoverpvsp1}). 
Pions were rejected by requiring a minimum energy of 0.06 GeV in the inner layer
of the calorimeter and a pulse height in the Cerenkov counter
corresponding to at least 2.5 photoelectrons
\cite{CC_nim_paper,EC_nim_paper}.

\begin{figure}[t]
\begin{center}
\includegraphics[height=6cm,clip=true,bb=1 1 540 380]{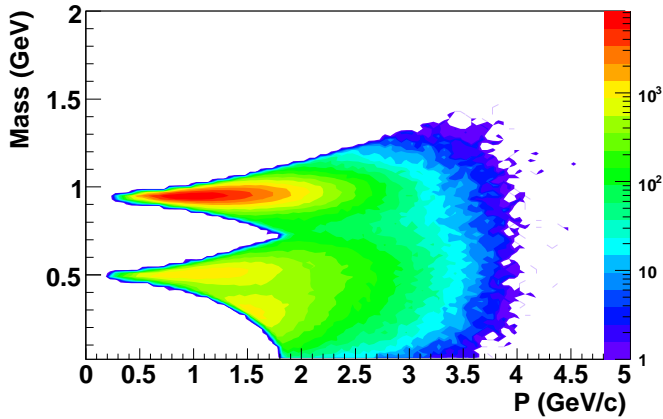}
\caption{\label{TOFMASS} Mass computed from the flight time versus momentum for positive particles. The top band corresponds to protons, the middle band corresponds to
$K^+$'s and the lower enhancement at 1.5 GeV/c momentum is due to 
pion contamination. }
\end{center}
\end{figure}

The two positive tracks in the fiducial volume were identified as a
proton and $K^+$ using the measured flight time ($\delta T \sim$ 160 ps) from the target to the
time-of-flight counters \cite{TOF_nim_paper}, a typical distance of about 5 m.  Fiducial
volume cuts were made to cut out tracks in inefficient
parts of the detector and small momentum corrections were applied to compensate
for uncertainties in the magnetic field and detector positioning.  The
time of the interaction was determined using the vertex time of the
electron corrected to the time of the bunch crossing of the machine.  
Using the known momenta of each of the
tracks, the vertex time was computed making assumptions for the mass
of the particle and comparing to the time of the bunch crossing. Events
were kept where the two positive tracks were consistent with the
assignment of one proton and one $K^+$. Tracks were identified as
protons when the projected vertex time assuming a proton mass differed from the interaction time
by less than 0.75 ns, and as a $K^+$ when the projected time assuming
a kaon mass differed from the interaction time by less than 0.6 ns. In cases where one track
satisfied both criteria, the ambiguity was resolved using the second
track. The number of events where both tracks satisfied both criteria
was less than 1\% and were eliminated. The calculated mass versus momentum, shown in Fig.\,\ref{TOFMASS}, 
indicates that at high momenta, there remain a number of pions that are identified as kaons in the sample.

\begin{figure}[t]
\includegraphics[height=6cm,clip=true]{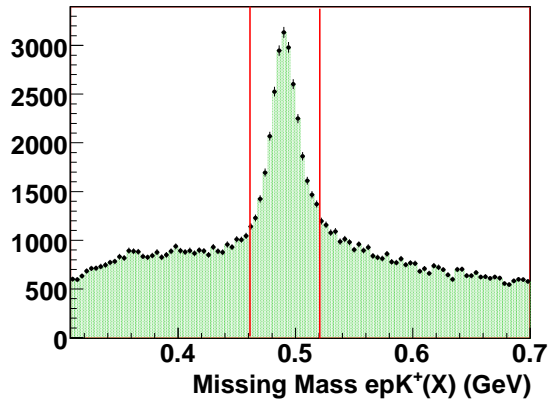}
\caption{\label{MMepkX}  (Color online) Distribution of $epK^{+}X$ missing mass. The vertical lines
indicate the cuts placed to select events with a missing $K^-$.}
\end{figure}

Once the electron, proton and $K^+$ tracks were identified, the missing mass was
computed and is plotted in Fig.\,\ref{MMepkX}. A clear peak is found at the mass
of the $K^-$, which corresponds to the exclusive reaction $e p \rightarrow epK^+K^-$. 
A $2\sigma$ cut was applied to the $epK^{+}X$ events to select the sample of
interest. For those events, the four-vector for the $K^-$ was constructed
by setting the three-momentum equal to the missing momentum of the $e p \rightarrow epK^+X$ reaction,  and the energy was then calculated
using the $K^-$ mass recommended by the Particle Data Group \cite{PDG}. The fraction of events where
the $K^-$ was detected in the detector was so small that they were not treated
differently than the rest of the sample.

\subsection{ $\phi$ Event Identification}

\begin{figure}[t]
\begin{center}
\includegraphics[height=6cm,clip=true,bb=4 4 565 385]{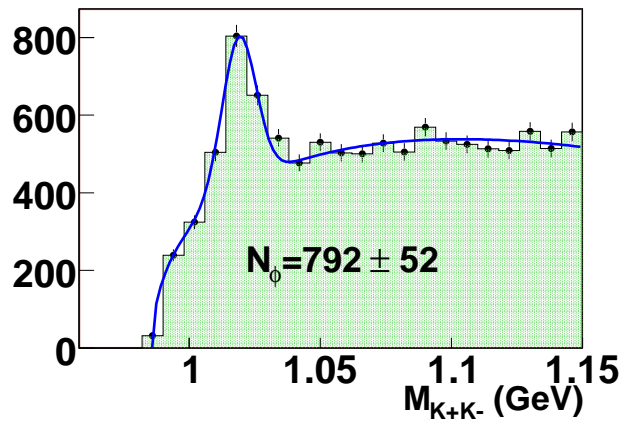}
\caption{\label{phifit} (Color online) $K^{+}K^{-}$ invariant mass including all data 
cuts and a fit to $\phi$ peak with Eq.\,\ref{fitfunction}.}
\end{center}
\end{figure}

The sample satisfying the $epK^{+}(K^{-})$ criteria contains 27,950
events out of 947,300 $epK^+X$ candidates. The sample includes all
physical processes that contribute to this final state, as well as real
$\phi$'s and background from misidentified pions.
Fig.\,\ref{phifit} shows the $K^{+}K^{-}$ invariant mass ($M_{KK}$) for the entire data set
with a clear $\phi$-meson peak. 
This distribution is
simultaneously fit to a Gaussian plus an empirical phase space function for the background, 
\begin{equation}\label{fitfunction}
\mathsf{FIT} =
A\mathsf{G(\sigma,\mu)}+B_{1}\sqrt{\mathsf{M_{kk}}^2-M_{th}^2}+B_{2}\Big(\mathsf{M_{kk}}^2-M_{th}^2\Big),
\end{equation} 
where $\mathsf{G(\sigma,\mu)}$ is a Gaussian distribution,
$M_{th}$=0.986 GeV is the threshold for two kaon production
and $A, B_{1},$ and $B_{2}$ are
parameters of the fit.  This fit yields $N_{\phi}=792\pm52$, a mean
$\mu=1.0194\pm 0.0005$ GeV, and a width of $\sigma=6.5\pm 0.6$ MeV. The signal-to-background
ratio for this fit was 0.56.
The mean and width are fixed to these values for all subsequent fits to the invariant mass
distribution to constrain the fits with limited statistics  in specific kinematic bins. A total 
of 37 distributions were fitted to extract the $\phi$ signal in various kinematic bins
(see subsequent sections for details).
The average $\chi^2$ per degree of freedom for all the fits was 1.07, indicating that deviations from 
the fit function are statistical in nature.

\begin{figure}[t]
\begin{center}
\includegraphics[height=6cm,clip=true,bb=6 258 263 427]{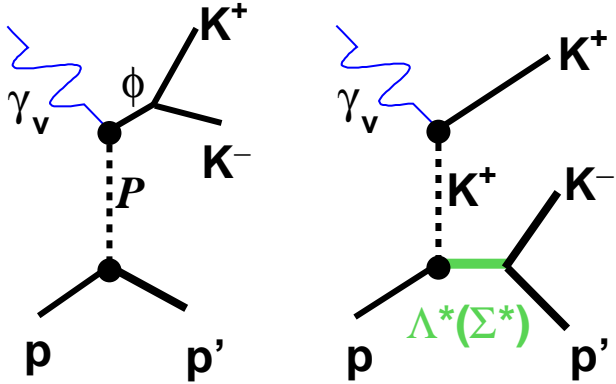}
\caption{\label{lambdafeyn} (Color online) Sketch of exclusive $\phi (1020)$ production via 
Pomeron exchange and of excited hyperon production,
of which $\Lambda (1520)$ is an example.
This is the primary physics background for $\phi (1020)$ production. }
\end{center}
\end{figure}

\begin{figure}[t] 
\includegraphics[height=5.5cm,clip=true,bb=15 15 600 350]{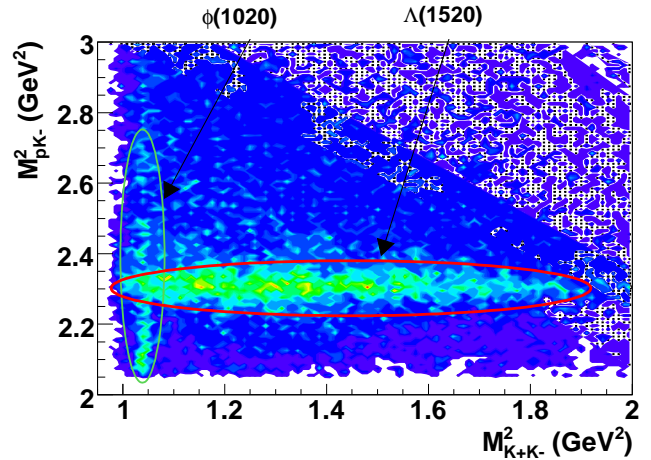}  
\caption{\label{imkk_vs_impk} Dalitz
plot of $M_{pK}^2$ versus $M_{KK}^2$. The well-defined horizontal
strip is the $\Lambda(1520)$ band. The vertical strip is the 
$\phi(1020)$ band.}
\end{figure}

There are competing physics channels that also lead to the same final
state. The majority of these backgrounds come from the production and
subsequent decay of high-mass hyperons produced via $ep\rightarrow
e^{'}K^{+}\Lambda^{*}(\Sigma^{*})$ as illustrated in Fig.\,\ref{lambdafeyn}.  
The Dalitz plot in Fig.\,\ref{imkk_vs_impk}
clearly shows the dominant $\Lambda(1520)$ background  contribution
(horizontal strip), as well as the $\phi(1020)$ (vertical strip). 
There are additional contributions from the higher-mass states such as
$\Lambda^{*}(1600)$, $\Lambda^{*}(1800)$, $\Lambda^{*}(1820)$,
$\Sigma^{*}(1660)$, and $\Sigma^{*}(1750)$ but they cannot be
separately identified. In order to avoid the introduction of holes in
the acceptance, no cuts are made to remove these hyperon
backgrounds. Instead they are taken into account during the fitting
procedure by assuming they contribute to the smooth background under the
$\phi$-meson peak. Nevertheless, many different fits were performed
removing events in the peak of the $\Lambda^*(1520)$ to study this
systematic with no indication that they changed the results
significantly. These studies focused on the t-distributions, since the
effective momentum transfer in $\Lambda^*$ reactions is very flat
compared to that expected from $\phi$-meson production.

\section{ Acceptance Corrections}

Particle interactions and event reconstruction in the detector were
simulated using a GEANT-based Monte Carlo called GSIM \cite{GSIM}.  The events
were generated according to a VDM-inspired cross section \cite{bauer} with the
following form:
\begin{eqnarray} 
\sigma_{\phi}^{VDM}(Q^2,W) & = & { \sigma_{\phi}(0,W) 
\left[1+R \epsilon \right] \over (1 + Q^2/M_{\phi}^2)^3 } \nonumber \\
& \times & {(W^2-M_{p}^2) \, \exp(-bt') \over \sqrt{(W^2-{M_{p}}^2-Q^2)^2+4W^2Q^2}} \\
\epsilon & = & {4 E_e (E_e - \nu) - Q^2 \over 4 E_e (E_e - \nu) + 2\nu^2 + Q^2},
\label{xsection}
\end{eqnarray}
where $\sigma_{\phi}(0,W)$ is the (transverse) photoproduction cross
section, $E_e$ is the incident electron beam energy, $\epsilon$ is the
virtual photon polarization parameter and $R$ is the ratio of the
longitudinal to transverse cross section.  The parameters of the
model were tuned during preliminary analysis and found to reproduce
the general features of the data. The main variation from the
conventional VDM model was in the propagator, where preliminary data
seemed to indicate a stronger dependence on $Q^2$ and an exponent of 3
was used instead of 2.

The acceptance function is
a combination of the geometrical acceptance of CLAS, the detector
efficiencies of the scintillators and drift chambers, the track
reconstruction efficiency, and the event selection efficiency. 
The Cerenkov detector \cite{CC_nim_paper} is not well modeled in GSIM,
and its efficiency was determined separately using the data.

\begin{figure}[t]
\begin{center}
\includegraphics[height=6cm,clip=true, bb=0 0 565 380]{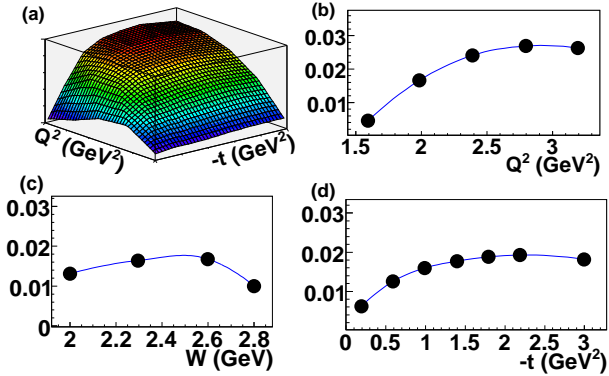}
\caption{\label{acc_q2_t} (Color online) 2-D Acceptance in $Q^2$ and $t$, as well as the 
1-D acceptance in $Q^2$, $W$, and $-t$. Error bars are not shown; the lines are present to guide the eye.
The axes in the 2-D plot in a) have the same range as that of the axis of the two 1-D plots in b) and d).
}
\end{center}
\end{figure}

\begin{figure}[t]
\begin{center}
\includegraphics[height=6cm,clip=true, bb=0 0 565 380]{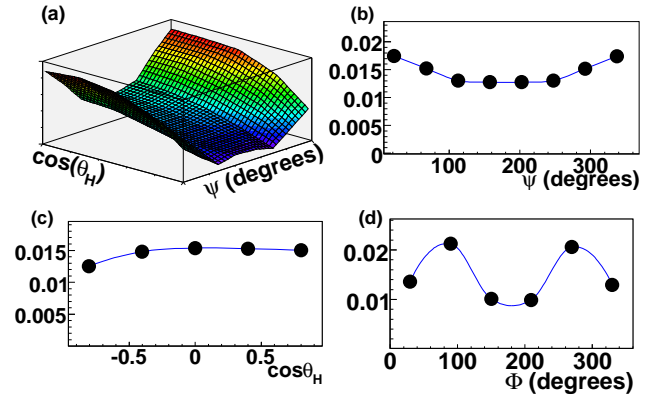}
\caption{\label{acc_costhetahpsiphi} (Color online) 2-D Acceptance in 
$\cos{\theta_{H}}$ and $\psi$, as well as the 
1-D acceptance in $\cos{\theta_{H}}$, $\Phi$, and $\psi$. 
Error bars are not shown;  the lines are present to guide the eye.
The axes in the 2-D plot in a) have the same range as that of the axis of the two 1-D plots in b) and c).
}
\end{center}
\end{figure}

The acceptance was defined in each bin of a 6-dimensional table as the
ratio of reconstructed to generated Monte Carlo events. In order to account for
correlations between all kinematic variables, a total of
33,600 acceptance bins are defined in the kinematic variables $Q^2$, $-t$, $W$,
$\Phi$, $\cos{\theta_H}$ and $\psi$. The binning selection is given in
Table\,\ref{binning} for the first three variables and uniform binning
was used for $\Phi$ (6 bins), $\cos{\theta_H}$ (5 bins) and $\psi$ (8
bins). The projected 2-D acceptance surface in $Q^2$ and $-t$ and the
1-D projections in $Q^2$, $t$, and $W$ are shown in
Fig.\,\ref{acc_q2_t} .  The projected 2-D acceptance surface in
$\cos{\theta_{H}}$ and $\psi$ is shown in Fig.\,\ref{acc_costhetahpsiphi},
as well as the 1-D projections in $\cos{\theta_{H}}$, $\psi$, and $\Phi$.
The variation of the acceptance is relatively smooth as a function of
these variables (except for $\Phi$, which is a reflection of the CLAS torus coils)
and is of the order of 1--3\%. 

Events that fell into bins with extremely small acceptances ($\leq$ 0.2\%)
were eliminated to avoid biases due to statistical fluctuations in those
bins. The losses were estimated and corrected by using the ratio of Monte Carlo 
acceptance-weighted events to generated events. 

\begin{table}[h]
\begin{center}
\begin{tabular}{|c|c|c|c|c|c|c|c|c|}
\hline
            & No. & \multicolumn{7}{|c|}{Bin Definition}                        \\ \hline
 {$Q^2$}    & 5       & 1.4-1.8 & 1.8-2.2 & 2.2-2.6 & 2.6-3.0 & 3.0-3.8  &         &         \\ \hline
 {$W$}      & 4       & 1.9-2.1 & 2.1-2.5 & 2.5-2.7 & 2.7-2.9 &          &         &         \\ \hline
 {$-t$}     & 6       & 0.4-0.8 & 0.8-1.2 & 1.2-1.6 & 1.6-2.0 & 2.0-2.4  & 2.4-3.6 &         \\ \hline
 {$t'$}     & 7       & 0.0-0.2 & 0.2-0.4 & 0.4-0.6 & 0.6-0.8 & 0.8-1.0  & 1.0-2.0 & 2.0-3.6 \\ \hline
\end{tabular}
\caption{Binning for the acceptance calculation in $Q^2$ (GeV$^2$), $-t$ (GeV$^2$), and $W$(GeV).
An additional acceptance table was also generated for $t'$ in the place of $t$, but it is not an independent variable.}
\label{binning}
\end{center}
\end{table}

\section{ Radiative Corrections}

\begin{figure}[t]
\begin{center}
\includegraphics[height=6cm,clip=true,bb=0 0 565 450]{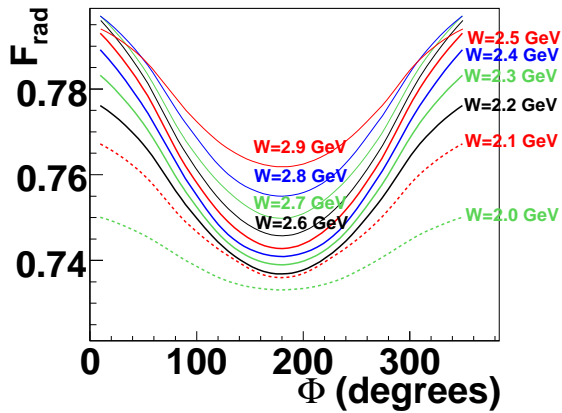}
\caption{\label{RADCOR}(Color online) Plot of radiative correction $F_{rad}$ as a 
function of $\Phi$ for assorted values of $W$ from 2.0 to 3.0 GeV.
The correction for each $W$ value was computed for $<Q^{2}>=2.47$ GeV$^2$ 
and $<\cos{\theta_{c.m.}}>=0.345$.}
\end{center}
\end{figure}

The radiative effects were calculated
in two distinct steps.  The external radiative process, which is the
finite probability that the incoming or scattered electron will
radiate a hard photon in the presence of a nucleon in the target other
than the one associated with the event, is taken into account during
the Monte Carlo acceptance calculation. The internal radiative
corrections include the Bremsstrahlung process for the incoming or
scattered electron in the presence of the nucleon associated with the
event, as well as diagrams such as vacuum polarization, which are not
accounted for during the acceptance calculation. These are
included in the correction factor $F_{rad}$
using the radiative correction code {\it EXCLURAD} setting the controlling parameter 
$v_{cut}=0.047$ GeV$^{2}$ \cite{RADCOR}.  $F_{rad}$  is calculated in each $W$ and $\Phi$ bin
as the ratio ${\sigma_{rad}/\sigma_{norad}}$ (variable $\delta$ in Eq.\,75 from Ref.\,\cite{RADCOR}), where $\sigma_{norad}$ is the cross
section calculated without any radiative effects (i.e. the Born
cross section) and $\sigma_{rad}$ is the cross section calculated with
radiative effects included.  The correction factor for various W bins
is shown as a function of $\Phi$ in Fig.\,\ref{RADCOR}. The
correction was computed in bins of $W$ and $\Phi$ for average values of $Q^2$
and $\cos{\theta_{CM}}$ (directly related to $-t$) because the correction was found to change
less than 2\% over the range of $Q^2$
and $\cos{\theta_{CM}}$\cite{radclasnote}.

\section{Cross Sections}
The reduced $\gamma^* p \rightarrow \phi p$ electroproduction cross section is given by
\begin{eqnarray}
\sigma(Q^2,W) & = & {1\over \Gamma(Q^2,W,E_{e})}{d\sigma \over dQ^2dW}, \nonumber \\
\Gamma(Q^2,W,E_{e}) & = & {\alpha \over 4 \pi} {W (W^2-M_p^2) \over M_p^2 E_e^2 Q^2} {1 \over 1-\epsilon},
\label{xsection1}
\end{eqnarray} 
where $\Gamma(Q^2,W,E_{e})$ is the virtual photon flux factor. 
We can extract the $\phi$ cross section from the data via
\begin{eqnarray}
{d\sigma \over dQ^2dW} & = & {1 \over Br} {n_{W}\over \mathcal{L}_{int} \Delta Q^2 \Delta W},
\label{xsection2}
\end{eqnarray}
where $Br$ is the branching fraction ratio of $\phi\rightarrow K^+K^- =0.491 \pm 0.009$ \cite{PDG},
$\mathcal{L}_{int}$=$2.98 \times 10^{40}$ cm$^{-2}$ is the live-time-corrected
integrated luminosity, $\Delta Q^2$ and $\Delta W$ are the corresponding
bin widths modified appropriately when not completely filled due to kinematics,
and $n_{W}$ is the result of a fit to the $M_{KK}$ distribution 
weighted by acceptance, CC efficiency correction and radiative corrections.
The binning in $Q^2$, $-t$, $t'$, and $W$ for the extraction of the cross section in those variables is shown in Table\,\ref{binning}. We emphasize here that we have performed a fit to Eq.\,\ref{fitfunction}
to determine the signal $n_{W}$ and the estimated background under the peak for each entry in the table.
The differential cross section in a variable $X$ is given as
\begin{equation}\label{diffxsectioneqn}{d\sigma \over dX} = {\sigma(Q^2)\over\Delta X },\end{equation}
where $\sigma(Q^2)$ is the total cross section in a bin in $X$.
The cross sections presented in this paper have been corrected
for the bin size and are quoted at the center of each bin.

\subsection{Total Cross Section $\sigma(Q^2)$}
The cross section $\sigma(Q^2)$ as a function of $Q^2$ is obtained
by integrating over $W$ due to the limited statistics.
Each event was weighted for acceptance,
radiative effects, the CC efficiency, as well as the virtual photon
flux factor.  The invariant mass distribution ($M_{KK}$) of weighted
events in each $Q^2$ bin was then fit to Eq.\,\ref{fitfunction}.
The bins used in the analysis
are given in Table\,\ref{kinematic_fid_cuts}. The range in $W$ was
restricted at the low end where acceptance corrections change
rapidly and are large, and at the high end to match the high end
of the kinematically accessible range.
\begin{table}[h]
\begin{center}
\begin{tabular}{|c|c|}
\hline
 {$Q^2$} range               & $W$ range                  \\ \hline \hline
  $ 1.4 \leq Q^2 \leq 1.8 $  & $2.10 \leq W \leq 2.90 $   \\ \hline
  $ 1.8 \leq Q^2 \leq 2.2 $  & $2.10 \leq W \leq 2.90 $   \\ \hline
  $ 2.2 \leq Q^2 \leq 2.6 $  & $2.10 \leq W \leq 2.90 $   \\ \hline
  $ 2.6 \leq Q^2 \leq 3.0 $  & $2.10 \leq W \leq 2.70 $   \\ \hline
  $ 3.0 \leq Q^2 \leq 3.8 $  & $2.10 \leq W \leq 2.70 $   \\ \hline
\end{tabular}
\caption{$W$ range for each $Q^2$ bin.}
\label{kinematic_fid_cuts}
\end{center}
\end{table}
The cross section for each of
the bins was calculated according to Eqs.\,\ref{xsection1} and
\ref{xsection2}. A small correction ($\sim$1-2\%) was then applied to
adjust the bin-averaged cross section to the center of the bin \cite{BIN_clasnote}. 
The values for the cross section in each $Q^2$ bin are shown in
Table\,\ref{TOTALXSECTION}. The total cross section was fit to the function
\begin{equation} A\over (Q^{2}+M_{\phi}^2)^{n} \end{equation}
to determine the scaling behavior. For this data we determined the parameter $n=1.97\pm 0.84$.
The measured exponent spans the range expected for the dependence on $Q^2$ due to
VDM ($n=2$) to hard scattering ($n=3$ for fixed momentum transfer $t$).

\begin{table}[h]
\begin{center}
\begin{tabular}{|c|c|c|c|}
\hline
   $Q^2$ (GeV$^2)$ &  $<\epsilon>$      & $\sigma$ (nb)          \\ \hline \hline
    1.6          &  0.488             &    9.9 $\pm$ 3.2       \\
    2.0          &  0.479             &   10.4 $\pm$ 2.5       \\
    2.4          &  0.471             &    6.7 $\pm$ 1.9       \\
    2.8          &  0.464             &    5.9 $\pm$ 2.4       \\
    3.4          &  0.452             &    3.6 $\pm$ 1.8       \\\hline
\end{tabular}
\caption{Total cross section $\sigma(Q^2)$ and kinematics of each data point, along 
with the center of each $Q^2$ bin. $<\epsilon>$ is the average virtual photon polarization
                in each bin. }
\label{TOTALXSECTION}
\end{center}
\end{table}

\subsection{Differential Cross Section in $t^{\prime}$, $d\sigma / dt^{\prime}$}
The differential cross section in $t^{\prime}$ was extracted in seven
bins in $t^{\prime}$ by fitting Eq.\,\ref{fitfunction}  to the $K^+K^-$ mass distribution to determine the 
$\phi$ signal and background in that particular bin. The average $\chi^2$
 per degree of freedom for these fits was 1.2.
The signal-to-background ratio varied from bin to bin, ranging from 0.33
to 0.86. The lowest signal-to-background ratio occurred in the mid range of $t'$. 
The resulting values for the cross section in each
$t^{\prime}$ bin are shown in Table\,\ref{DIFFXSECTION2}.  

\begin{figure}[t]
\begin{center}
\includegraphics[height=6cm,clip=true,bb=1 1 560 536]{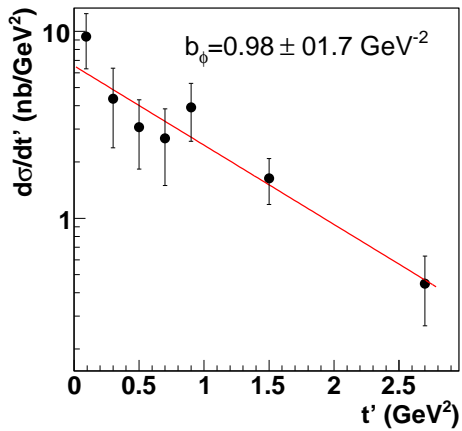}
\caption{\label{fulltprimefit}(Color online) Plot of $d\sigma/dt^{\prime}$ along with 
an exponential fit.}
\end{center}
\end{figure}

In cases of limited statistics, $d\sigma / dt^{\prime}$ is often used
instead of  $d\sigma / dt$ in order
to eliminate kinematic corrections due to $-t_{0}$, which varies with $Q^2$ and $W$.
This procedure is most useful when the cross section factorizes into terms
that depend only on $t$ and terms that depend on $Q^2$ and $W$, aside from
the threshold dependence, as in the VDM model. Indeed, our measured
differential cross section in $t'$ show very similar trends as
previous data, namely they are consistent with diffractive production
($e^{-b_{\phi}|t'|})$ \cite{costypaper}. Fig.\,\ref{fulltprimefit} 
shows an exponential fit to the measured differential cross
section, which yields a $b_{\phi}=0.98 \pm 0.17$ GeV$^{-2}$. At high
energies, the slope can be directly interpreted in terms of the
transverse size of the interacting configuration, as described 
later when presenting results. In that limit, the
small value of the exponential slope implies the interaction
takes place at very short distances inside the nucleon.

\begin{table}[h]
\begin{center}
\begin{tabular}{|c|c|c|}
\hline
  $ t^{\prime} $ (GeV$^2)$ & $d\sigma/dt^{\prime}$ (nb/GeV$^2$)  \\ \hline
    0.1               &  9.4  $\pm$ 2.9       \\
    0.3               &  4.4  $\pm$ 1.9       \\
    0.5               &  3.1  $\pm$ 1.2       \\
    0.7               &  2.7  $\pm$ 1.2       \\
    0.9               &  4.0  $\pm$ 1.3       \\
    1.5               &  1.6  $\pm$ 0.4       \\
    2.8               &  0.5  $\pm$ 0.2       \\ \hline
\end{tabular}
\caption{Differential cross section $d\sigma/dt^{\prime}$ and 
kinematics of each data point. $t^{\prime}$ 
is the center of the bin, and corresponds to an average value of 
$\epsilon=0.47$. }
\label{DIFFXSECTION2}
\end{center}
\end{table}

\subsection{Differential Cross Section in $t$, $d\sigma / dt$}
The differential cross section is easiest to compare with theory if it
is computed in terms of the Mandelstam variable $t$.  The cross
section is given as
\begin{eqnarray} 
{d\sigma \over dt} & = &
{\sigma(Q^2)\over\Delta t \cdot \mathsf{Corr}(t_{0})}, 
\end{eqnarray}
where $\Delta t$ is the bin size and
$\mathsf{Corr}(t_{0})$ is a correction factor to account for the fact
that the kinematic limit $t_{0}(Q^2,W)$ varies across the bin.
The yield was extracted over the ranges of $Q^2$ and $W$ 
given in Table\,\ref{kinematic_fid_cuts} in six bins
in $-t$.
The kinematic threshold $t_0$ varies between -0.09 and -1.14 GeV$^2$
for extreme values of $Q^2$ and $W$. For the bin corresponding to
$0 \leq -t \leq 0.4$ GeV$^2$, the threshold varies so much that
corrections could not be modeled
reliably, so that bin was dropped. The first bin reported contains
a significant correction, but was included with an increased
systematic error. Subsequent bins had small or no corrections. 
The values for the cross section in each $-t$ bin are given in
Table\,\ref{DIFFXSECTION}.

\begin{table}[h]
\begin{center}
\begin{tabular}{|c|c|c|}
\hline
  $-t$ (GeV$^2)$ & ${d\sigma/dt}$ (${nb/ GeV^2}$)   \\ \hline
   0.6      &  10.7  $\pm$   3.1        \\
   1.0      &   0.8  $\pm$   1.0        \\
   1.4      &   3.4  $\pm$   1.0        \\
   1.8      &   1.0  $\pm$   0.5        \\
   2.2      &   1.4  $\pm$   0.5        \\
   3.0      &   0.5  $\pm$   0.2        \\ \hline
\end{tabular}
\caption{Differential cross section $d\sigma/dt$ and kinematics 
of each data point. $-t$ is the center of each bin at an average value of
$\epsilon=0.47$.}
\label{DIFFXSECTION}
\end{center}
\end{table}

\subsection{Differential Cross Section $d\sigma / d\Phi$ and test of SCHC}
The cross section dependence on the angle $\Phi$ between the electron
and hadron scattering planes takes the following form:
\begin{eqnarray}
{d\sigma\over d\Phi} & = & {1\over 2\pi}\Bigg(\sigma+\epsilon
\sigma_{TT}\cos 2\Phi+\sqrt{2\epsilon(1+\epsilon)}\sigma_{LT}\cos\Phi\Bigg), \nonumber \\
 &\, & 
\label{Wphi2}
\end{eqnarray} 
where $\sigma_{LT}$ and $\sigma_{TT}$ are the interference terms between the 
longitudinal and transverse contributions to the cross section.
If helicity is conserved in the s-channel (SCHC), then both of these terms
will vanish. The magnitude of these interference terms can therefore
be used as a test for the validity of SCHC.

\begin{figure}[t]
\begin{center}
\includegraphics[height=6cm,clip=true,bb=1 1 565 530]{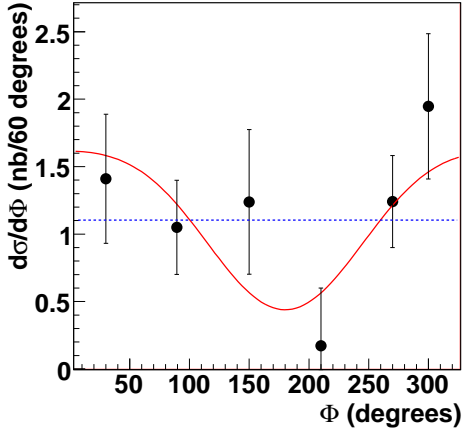}
\caption{\label{dsigdphi} (Color online) $d\sigma/d\Phi$ vs $\Phi$. The curve 
shows a fit to Eq.\,\ref{Wphi2} which is used to determine
$\sigma_{TT}$ and $\sigma_{LT}$. The
dotted line is a fit to a constant function which is expected from 
SCHC.}
\end{center}
\end{figure}

The differential cross sections in $\Phi$ were extracted in the same
manner as the other differential cross sections (Eq.\,\ref{diffxsectioneqn})
after integrating over $Q^2$,
$-t$ and $W$. The cross section  $d\sigma/d\Phi$
was extracted in six bins in $\Phi$.
The cross sections, along with a fit to Eq.\,\ref{Wphi2}, are shown in
Fig.\,\ref{dsigdphi}.  The fit yields a
value of 
$\sigma_{TT}=-1.1 \pm 3.1$~nb and $\sigma_{LT}=2.2 \pm 1.1$~nb
with a chi-squared per degree-of-freedom of  $\chi^{2}/D.F.=1.3$. A fit of the $d\sigma/d\Phi$ distribution to a constant,
constraining the interference terms to be zero, yields a
$\chi^{2}/D.F.=1.6$. The small change in the goodness of fit
between the two cases leads us to conclude that the precision of this experiment is insensitive to
violations of  SCHC for $\phi$-meson production in our kinematic domain.

\section{Angular Decay Distributions}
The angular decay distribution of the $K^{+}$ in the $\phi$
rest frame describes the polarization properties of the $\phi$-meson.
The scattering amplitude for vector meson
electroproduction ${\gamma^{*}+ N \rightarrow P + V}$ can be expressed
in terms of the helicity amplitudes
$T_{\lambda_{V}\lambda_{P}\lambda_{\gamma^{*}}\lambda_{N}}$, where
$\lambda_{i}$ is the helicity of each particle
(i=V, P, $\gamma$, N).  The vector meson spin
density matrix is derived from these helicity amplitudes by exploiting
the von Neumann formula 
\begin{equation}
\rho(V)= {1\over 2}T\rho(\gamma^{*})T^{\dagger},
\end{equation} 
where $\rho(\gamma^{*})$ is the spin-density matrix of the virtual photon.  
The details of this derivation can be found in Ref.\,\cite{wolfShilling}.
The density matrix element is denoted $\rho_{ij}^{\alpha}$, where the index 
$\alpha$ can be related to the virtual photon polarization. $\alpha=0-2$ 
for purely transverse photons, $\alpha=4$ for purely longitudinal photons, 
while other values correspond to longitudinal-transverse interference terms. 
The indices $ij$ correspond to the helicity state of the vector meson
\cite{pomeron}.  In cases where the data do not allow for a
$\sigma_{L}/\sigma_{T}$ separation, the unseparated matrix 
elements $r_{ij}^{\alpha}$ can be parameterized as:
\begin{eqnarray} 
r_{ij}^{04} = {\rho_{ij}^{0}+\epsilon R\rho_{ij}^{4}\over {1+\epsilon R}} & &   \\ 
r_{ij}^{\alpha} = {\rho_{ij}^{\alpha} \over {1+\epsilon R}} &;& \alpha = 0-3              \\ 
r_{ij}^{\alpha} = \sqrt{R} {\rho_{ij}^{\alpha}\over {1+\epsilon R}} &;& \alpha = 5-8, 
\end{eqnarray}
Recall that $R$ is the ratio of longitudinal to transverse cross section.
The angular distribution of the $K^+$ is usually
described in the helicity frame, defined in the rest frame 
of the $\phi$-meson with the z-axis oriented along the
$\phi$-meson in the $\gamma^* p$ center-of-mass. The full
decay distribution, which we denote by  $\mathsf{W_{F}}(\cos{\theta_{H}},\Phi,\phi_{H})$,
can be found in the literature \cite{ShillingSeybo},
but will only be given here in simplified forms. In particular, further analysis of angular
distributions is done under the assumption of SCHC, which 
leads to considerable simplifications with the introduction 
of $\psi=\phi_{H}-\Phi$ and the following constraints:
\begin{eqnarray}
{\mathsf{-Im}r^{6}_{10} = \mathsf{Re}r{^5}_{10}} 
& = &{{\sqrt{R}\cos{\delta}} \over {\sqrt{8}(1+\epsilon R)}} \label{SCHC1}; \\
r^{1}_{1-1} = -\mathsf{Im}r{^2}_{1-1} & = & {{1\over 2(1+\epsilon R)}} \label{SCHC2}; \\
r_{00}^{04} & = & {\epsilon R\over 1+\epsilon R} \label{SCHC4}.
\end{eqnarray} 
All other $r_{ij}^{\alpha}$'s are 0, and $\sqrt{R} e^{i\delta}$ is the ratio
of the longitudinal to transverse amplitudes. The angular
distribution becomes a function of two variables only and is given by:
\begin{eqnarray}
&& W(\cos\theta_{H},\psi) = {3\over 8\pi}{1\over 1+\epsilon R}\Bigg[\sin^2\theta_{H}+2\epsilon R\cos^2\theta_{H} \nonumber  \\
&& \hspace{0.5cm}-\;2(1+\epsilon R)\epsilon(r_{1-1}^{1})\sin^2\theta_{H}\cos2\psi \nonumber \\
&& \hspace{0.5cm}+\;4(1+\epsilon R)\:\sqrt{\epsilon(1+\epsilon)}(\mathsf{Re}r_{10}^{5})\sin2\theta_{H}\cos\psi\Big].
\label{SCHCWDIST}
\end{eqnarray}

In order to extract the $r_{ij}^{\alpha}$ parameters from
the measured angular distribution, we use two
1-dimensional projections of the full angular distribution.

\subsection{Polar Angular distribution projection}
\begin{figure}[t]
\begin{center}
\includegraphics[height=6cm,clip=true,bb=1 1 565 538]{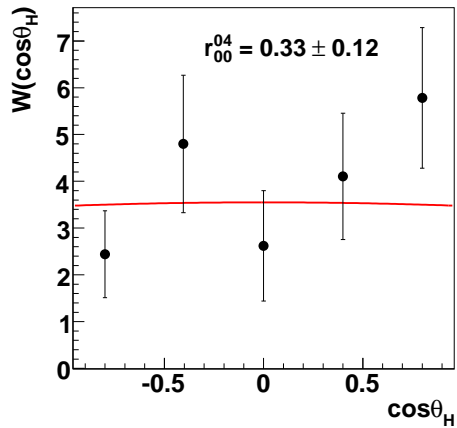}
\caption{\label{dN/dcosthetaH} (Color online) Unnormalized polar angular decay distribution of
the $K^+$ integrated over all $Q^2$ values plus a fit to Eq.\,\ref{costhetaproj}. 
Also shown is the extracted $r_{00}^{04}$ parameter.}
\end{center}
\end{figure}

To obtain the polar angular distribution, an integration of the full
angular distribution $\mathsf{W_{F}}$ over $\phi_{H}$ yields
\begin{equation}
W(\cos\theta_{H})={3\over 4}\Bigg[\Big(1-r_{00}^{04}\Big)+\Big(3r_{00}^{04}-1\Big)\cos^2\theta_{H}\Bigg],
\label{costhetaproj}
\end{equation} 
which is independent of SCHC. In order to
obtain this projection from the data, 
the $K^{+}K^{-}$ invariant mass distribution is plotted in five bins in $\cos{\theta_{H}}$
(0.40 units of $\cos{\theta_{H}}$ each). The same fit to a Gaussian plus
a polynomial background was made to extract the weighted yields in
each of these bins.
The fit to ${d\sigma/d\cos{\theta_{H}}}$ in Fig.\,\ref{dN/dcosthetaH}
yields a value $r_{00}^{04}=0.33 \pm 0.12$ with a $\chi^{2}/D.F.=1.7$.
With the additional assumption of SCHC, this parameter can be used to
determine the ratio of the longitudinal to transverse cross sections as
\begin{eqnarray}
R & = & {{r_{00}^{04}}\over {\epsilon (1-r_{00}^{04})}} = 1.05 \pm 0.38,
\end{eqnarray}
where we have used the average value of $<\epsilon>=0.47$.


\subsection{Angular distribution projection in $\psi$}
\begin{figure}[t]
\begin{center}
\includegraphics[height=6cm,clip=true,bb=1 1 566 530]{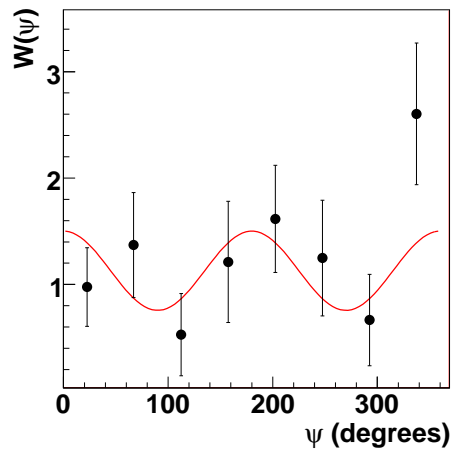}
\caption{\label{dN/dpsi} (Color online) Unnormalized azimuthal angular 
distribution extracted for all $Q^2$ values plus 
a fit to Eq.\,\ref{psiproj}. The value of $r_{1-1}^{1}$ can also
be used to determine $R$.}
\end{center}
\end{figure}

After an integration of $\mathsf{W_{F}}$ in $\cos{\theta_{H}}$, a
substitution of $\phi_{H}=\psi+\Phi$, and an integration in $\Phi$,
the projected angular distribution in $\psi$ is given as
\begin{equation}
W(\psi)={1\over
2\pi}\Bigg[1+2\epsilon(r_{1-1}^{1})\cos2\psi\Bigg],
\label{psiproj}
\end{equation}
which assumes SCHC.  The factor of $1/2\pi$ is a normalization factor. 
A fit of $d\sigma/d\psi$ to Eq.\,\ref{psiproj} is shown in Fig.\,\ref{dN/dpsi}.  The fit yields a value 
$r_{1-1}^{1}=0.38 \pm 0.23$
with a $\chi^{2}/D.F.=1.3$. The ratio
of longitudinal to transverse cross sections can also be computed from
$r_{1-1}^{1}$ (Eq.\,\ref{SCHC2}) and gives $R=0.72 \pm 0.3$, in agreement with the value
obtained previously.


\section{Systematic Uncertainties}
The relatively low number of measured $\phi$ events causes statistical errors to dominate.  
The sources of systematic errors in this experiment are summarized in 
Table\,\ref{syserror}.  The major sources of systematic errors are due
to acceptance corrections and estimation of backgrounds. Studies of
backgrounds and their uncertainties were also limited by the finite
sample size. The total systematic error of 18.6\% was added in 
quadrature with the statistical errors in all quoted cross sections.

The acceptance correction contributes to the systematic error in two
distinct ways. The uncertainty of 6\% introduced by eliminating events
with very large weights (i.e.\,very low acceptance) was estimated by changing the maximum
weight allowed and recomputing the extracted cross section. The
uncertainties introduced by the use of our acceptance table (12\%)
were estimated by combining bins and comparing the extracted result to
the average of the constituent bins.

To estimate the systematic uncertainty due to the unknown distribution
of backgrounds, the functional form of the background (see Eq.\,\ref{fitfunction}) 
was modified by adding a term proportional to
$(M_{KK}^3 - M_{th}^3)$ and refitting the $-t$ and $Q^2$
distributions. The new fits were less constrained, but the average
change in cross section was 9\%. We found that the extraction of the slope 
parameter $b_{\phi}$ was fairly robust to these changes.
In addition, the fitted invariant
mass distributions included some background due to 
misidentified pions. The estimated uncertainty due to this contamination
under the peak was estimated to be 7\%.

The systematic uncertainty in the placement of the cut to select
the $K^-$ from the $epK^+$ missing mass (5\%) was investigated by varying
the cut and observing the effect on the cross sections.  The systematic
error associated with the bin centering correction is almost negligible
($\sim 1$\%). The contribution to the systematic error from the
radiative correction was estimated to be $\sim 3$ \% and is described
in more detail in Ref. \cite{RADCOR}.  The fluctuation in the number of
photoelectrons in the Cerenkov counter over the course of the run can
cause a systematic error in the Cerenkov counter efficiency
correction. This leads to a systematic of $\leq 1\%$. Finally, 
the procedure to estimate the correction due to the $t_{0}$ 
kinematic cutoff in the first $t$-bin introduces a 25\% systematic error in that bin.

\begin{table}[h]
\begin{center}
\begin{tabular}{|l|c|}
\hline
  Source                             &    $\Delta \sigma$ \% \\ \hline \hline 
  Acceptance correction              &    13.4               \\  
  Background functional form         &    9.1                \\ 
  Misidentified pion background      &    7.0                \\
  $epK^+(X)$ cut                     &    5.0                \\
  Bin-centering correction           &    1.0                \\ 
  Radiative correction               &    3.0                \\ 
  Cerenkov efficiency correction     &    1.0                \\ \hline
  Total                              &   18.6                \\ \hline
\end{tabular}
\caption{Table of systematic errors.}
\label{syserror}
\end{center}
\end{table}

\section{Discussion}
\begin{figure}[t]
\begin{center}
\includegraphics[height=6cm,clip=true,bb=1 1 565 540]{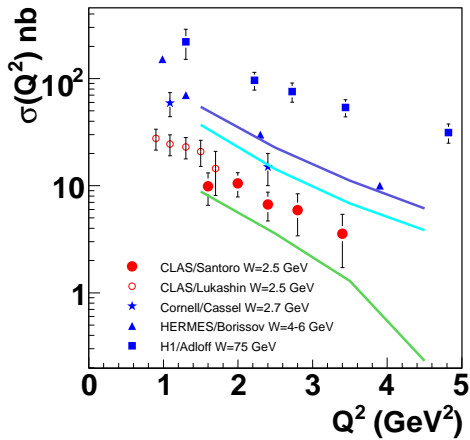}
\caption{\label{allq2datacomp} (Color online) Total cross sections as a function of $Q^2$ for our 
data (red full circles), previous JLab data (open circles) \cite{costypaper},
Cornell data (stars) for $W$ between 2 and 3.7 GeV \cite{Cassel}, 
HERMES data (triangles) for $W$ between 4 and 6 GeV \cite{Borissov}, and HERA data (squares) at high $W$ \cite{ZEUS_HERA_2005}.
The curves show the predictions of the JML model at $W$=2.9, 2.45 and 2.1 GeV (top to bottom).
}
\end{center}
\end{figure}

The measurements of $\sigma(Q^2)$ from the present analysis are
shown along with other data on $\phi$ electroproduction 
\cite{costypaper,Dixon2,Cassel,Borissov,ZEUS_HERA_2005} in Fig.\,\ref{allq2datacomp}. 
The one overlap point at $Q^2=1.5$ $GeV^2$ is in good agreement with the previous CLAS 
measurement \cite{costypaper}. The data sets span the range from threshold 
at $W$=2 GeV up to HERA energies. 

The data sets have a similar trend as a function of $Q^2$ and increase
monotonically as a function of $W$. The three curves using the JML
model at $W$ = 2.1, 2.45 and 2.9 GeV are also plotted for
Q$^2$ greater than 1.5 GeV$^2$.
The calculation for $W$=2.45 GeV, which is close to the average of our
data, seems to overestimate our data by about a factor of two, although
it does reproduce the existing Cornell data from Ref.\,\cite{Cassel}. 
The Cornell data has a much wider acceptance range in
$W$ between 2.0 and 3.7 GeV, so in fact it could be representative of the
cross section at higher $W$. The new data from CLAS, together
with the existing world data, in particular the data from HERA, indicate that the
qualitative behavior as a function of $Q^2$ does not change between
threshold and a $W$ of about 100 GeV.

\begin{figure}[t]
\begin{center}
\includegraphics[height=6cm,clip=true,bb=1 1 490 350]{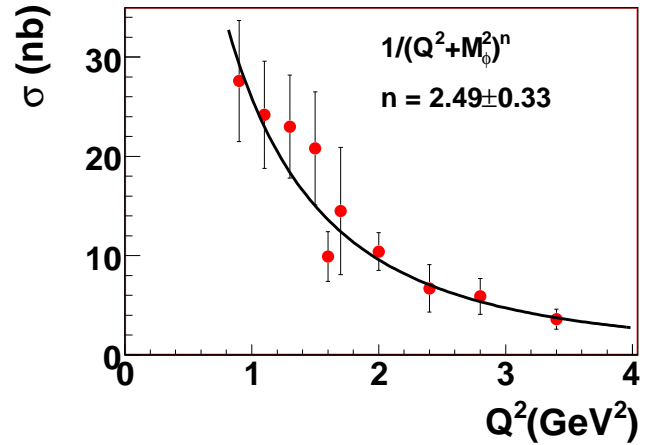}
\caption{\label{q2scaling} (Color online) Fit to the cross section as
a function of $Q^2$ distribution to determine scaling
using data from the present experiment and CLAS data from Ref.\,\cite{costypaper}.}
\end{center}
\end{figure}

Of interest is the applicability of factorization and the formalism of
GPDs to meson production in general, and $\phi$ production in
particular. QCD factorization makes certain asymptotic predictions
about the cross section, namely that the longitudinal part of the
cross section, $\sigma_{L}$, becomes dominant as $Q^2$ increases, and
that the differential cross section will scale as $1/(Q^2)^3$ at fixed
$t$ and $x_B$. 
For a slow variation of the cross section over the
range of $x_B$ of the data (0.2--0.5), this prediction can be compared to
the $Q^2$ dependence integrated over $W$ and $t$, although quantitative estimates are 
modified by power corrections as well as kinematics near threshold.
On the other hand, 
the VDM model predicts the cross section to scale as
$1/(Q^2+M_{\phi}^2)^n$ with $n=2$.  The Q$^2$ range of our data is
limited, but in combination with previous CLAS data at lower Q$^2$
\cite{costypaper} (see Fig.\,\ref{q2scaling}) we can determine the
scaling exponent of $1/(Q^2+M_{\phi}^2)^n$ to be $n=2.49\pm 0.33$.

Present theoretical calculations of the $\phi$ production cross section 
based on GPD models suffer from considerable quantitative uncertainties
when applied to fixed--target energies. At HERA energies the approach taken
in Ref.\,\cite{Brodsky}, which relies on the equivalence of 
leading-order QCD factorization with the dipole picture of high--energy scattering, 
gives a good description of the absolute cross section, as well as of 
subtle features such as the change of the $W$-- and $t$--dependence 
with $Q^2$. Essential for the success of this 
approach is the fact that the effective scale of the gluon GPD, 
$Q^2_{\text{eff}}$, is considerably smaller than the external photon 
virtuality, $Q^2$, as has been confirmed by detailed quantitative 
studies \cite{Frankfurt}. The same is expected in vector meson 
production at fixed--target energies; however, implementing it in a 
consistent manner in these kinematics has so far proven to be difficult. 
Leading-twist, leading-order QCD calculations of the $\phi$ production cross 
section at JLab and HERMES energies done with the assumption that
$Q^2_{\text{eff}} = Q^2$ \cite{Diehl:2005gn} overestimate the 
measured cross section by a factor 5--10 and predict too steep an energy 
dependence. A satisfactory solution to this problem likely requires a 
comprehensive approach that combines contributions from small--size 
($\sim 1/Q$) and hadronic--size configurations in the virtual photon
at moderate coherence lengths ($c\tau \lesssim 1 \, \text{fm}$),
and possibly higher--order (NLO) QCD corrections. 
We note that a modified perturbative approach \cite{Goloskokov:2006hr} which
includes the intrinsic transverse momentum in the meson wave function has
been fairly successful in reproducing the measured cross sections down to relatively low
$Q^2$ and $W$.

\begin{figure}[t]
\begin{center}
\includegraphics[height=6cm,clip=true,bb=1 1 565 535]{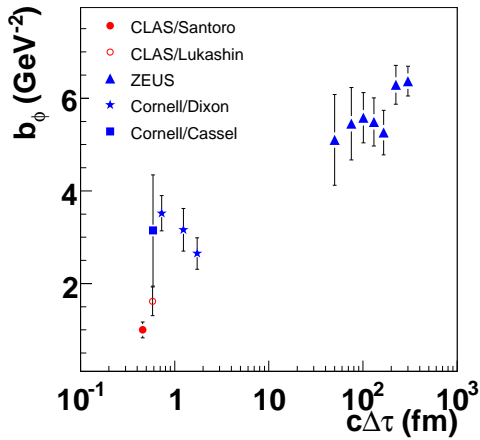}
\caption{\label{bphi} (Color online) Exponential slope $b_{\phi}$  plotted as a function of the
fluctuation parameters $c\Delta \tau$ for the world data. The data at high $W$ measure an asymptotic
slope corresponding to long fluctuation times. At low $W$ and relatively large $Q^2$,
the fluctuation times becomes small and constrain the size of the interaction volume.}
\end{center}
\end{figure}

The four-momentum transfer distribution probes the size of the interaction volume.
At high energies, the
exponential slope (see Fig.\ref{fulltprimefit}) is directly
related to the transverse size
$b_{\phi} \sim \frac{1}{3}R_{int}^2$ in analogy to the classical scattering of
light through an aperture of radius $R_{int}\sim$ 0.38 fm.
At energies close to threshold, as in the present experiment,
this interpretation requires some modification. 
When the coherence length becomes comparable to
the size of the target, longitudinal shrinkage occurs and this 
also causes a decrease of the exponential slope.
The longitudinal size is related to the fluctuation time $\Delta\tau$ 
of the virtual meson, which can be estimated through
uncertainty principle arguments, and is given by  Eq.\,\ref{DelTau}.
The nature of the interaction becomes more point-like as $Q^{2}$
increases and the fluctuation time decreases. This transition should
be observed as a decrease in the measured slope parameter.  Since the
differential cross section in $t^{\prime}$ was extracted for all
$Q^2$, the value for $b_{\phi}$ corresponds to the average value of
$c\Delta\tau$=0.46 fm. The slope parameters for various experiments are
shown in Fig.\,\ref{bphi} for the world data on $\phi$ electroproduction.  
The measured slower fall-off  of the $t$-distribution, corresponding to the small slope parameter, 
is consistent with the expectation that short
interaction time probe small $s \overline{s}$ dipoles. 

\begin{figure}[t]
\begin{center}
\includegraphics[height=6cm,clip=true,bb=3 3 560 531]{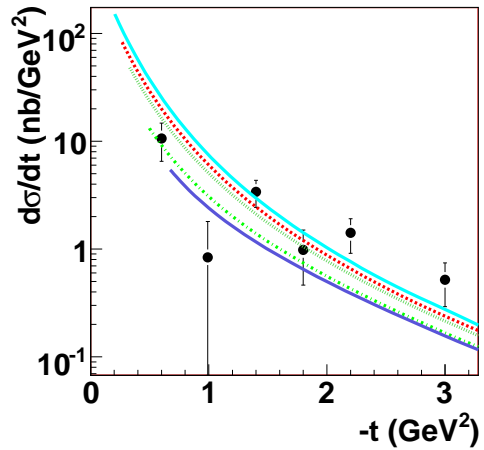}
\caption{\label{dsigdtallq2} (Color online) $d\sigma/dt$ vs $-t$ for 
the entire $Q^2$ range and the JML predictions for $W$=2.5 GeV 
at five values of $Q^2$= 1.6, 2.1, 2.6, 3.8 and 5 GeV$^2$, top to bottom.}
\end{center}
\end{figure}

The differential cross section in $-t$ is compared to the JML model in
Fig.\,\ref{dsigdtallq2}.  The data covers $1.4\leq Q^2 \leq 3.8$
GeV$^2$ and the JML model predictions \cite{JMLagetPrivate} are
plotted for fixed values of $Q^2$ from 1.6 to 5 GeV$^2$. The data tend
to have a shallower slope than the calculation, but there is general
agreement. This agreement is highly non-trivial since the few parameters of
the model have been fixed at the real photon point and kept frozen in the
virtual photon sector. Our data confirm both the $Q^2$ and $-t$ dependence
of the cross section that are built into the dynamics of the $s\overline{s}$
loop and the 2-gluon loop. 

The angular decay distributions provide information on the
longitudinal part of the production cross section.
We have extracted values of $\sigma_{TT}$ and $\sigma_{LT}$
from the cross section dependence on the angle $\Phi$ between
the electron and hadron scattering planes. The value of the $\sigma_{LT}$
is consistent with zero and the 
assumption that SCHC is valid for $\phi$ production in this kinematic regime.
However, small deviations are still possible as shown
by more accurate
measurements of these parameters at HERA energies \cite{Adloff1}.

The ratio $R={\sigma_{L}/\sigma_{T}}$ has been determined from two
projections of the angular decay distribution of the $K^+$ in the
$\phi$-meson rest frame and under the hypothesis of SCHC. The
measurement of $r_{00}^{04}$ gives $R=1.05 \pm 0.38$ and the
measurement of $r_{1-1}^{1}$ gives a value of $R=0.72 \pm 0.30$, the
weighted average being $R=0.85 \pm 0.24$. This measurement can be compared to
the value of R=1.25 predicted by the JML 2-gluon exchange model.  We
note that these extractions, at least from $r_{00}^{04}$, are
relatively insensitive to the assumption of SCHC as shown in
Ref.\,\cite{Adloff1}.

\begin{figure}[t]
\begin{center}
\includegraphics[height=6cm,clip=true,bb=1 1 565 540]{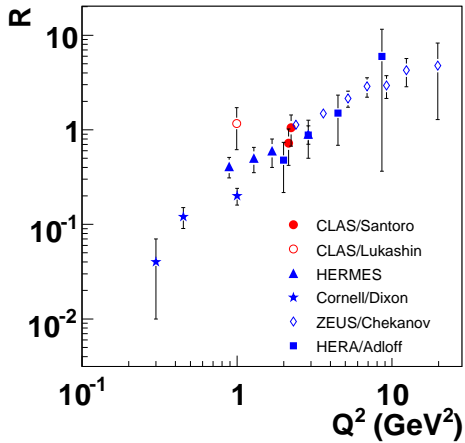}
\caption{\label{Rvsq2} (Color online) $R=\sigma_{L}/\sigma_{T}$ vs. $Q^2$ for 
our data (solid circles), previous CLAS result (open circle), HERMES results (triangles)
Cornell data (stars), ZEUS data (open diamonds) and HERA data (squares). The two
determinations from the present analysis 
are separated for ease of viewing about the actual $Q^2$ value of 2.21 GeV$^2$.}
\end{center}
\end{figure}

The measurements of $R$ from this analysis and other world data are plotted as
a function of $Q^2$ in Fig.\,\ref{Rvsq2} \footnote{The $W$-dependence
of $R$ has been studied at HERA \cite{ZEUS_HERA_2005}, which covers 
a very large range in $W$.}.  The data show that the ratio $R$ is
increasing as a function of $Q^2$, but $\sigma_{L}$ is still not
dominant at these kinematics.  Using our measurement of $R$, we can
compute the average longitudinal cross section for our data.  The
average cross section is given by $\sigma(Q^2=2.21 GeV^{2})=6.9 \pm 1.7$ nb,
which yields a longitudinal cross section $\sigma_L=4.5 \pm 1.1$~nb.
 
\section{Summary}
$\phi$-meson electroproduction was examined in the kinematical regime
$1.4\leq Q^2\leq3.8$ GeV$^2$, $0.0\leq t^{\prime}\leq 3.6$ GeV$^2$, and $2.0\leq W\leq
3.0$ GeV. This data set doubles the range of $Q^2$ previously reported at
JLab energies \cite{costypaper}, accruing approximately four times the
luminosity required for sensitivity to smaller cross sections. We have
presented distributions as a function of the momentum transfer $-t$,
the azimuthal angle $\Phi$ between the electron and hadron scattering
planes, as well as angular decay distributions in the rest frame of
the $\phi$-meson.  

We have analyzed the angular distributions under the assumption of
SCHC to extract the  ratio of longitudinal to transverse cross sections of
$R=0.85 \pm 0.24$, which is consistent with the world trend. The 
longitudinal component is comparable to the transverse one, 
which suggests that we have not yet reached the asymptotic regime
where QCD factorization can be applied without substantial corrections.


The cross sections have a weak dependence on $-t$,
which indicates that at this $Q^2$, the photons couple to 
configurations of substantially smaller size than the target.
Our data provide a very precise measurement 
of the exponential slope $b_{\phi}$ at
small $c \Delta \tau \sim$ 0.5 fm, which shows that we are probing
very small distances, approaching about one third the size of the
proton itself.
A natural explanation is that $\phi$
production is dominated by the scattering of small size 
$s\overline{s}$ virtual pairs off the target.  This conclusion is
supported by the good agreement between our data and the extension of
the JML model from the real photon point (where it has been
calibrated) to the virtual photon sector. It describes the interaction
between this $s\overline{s}$ pair and the nucleon by the exchange of
two dressed gluons. We conclude that these constituent degrees of freedom
are appropriate for the description of $\phi$-meson production at
low $W$ and $Q^2 \sim$ 2-3 GeV$^2$.

\section{Acknowledgments}

We would like to acknowledge the outstanding efforts of the staff of
the Accelerator and the Physics Divisions at JLab that made this
experiment possible. 
This work was supported in part by 
the U.S. Department of Energy, 
the National Science Foundation, 
the Italian Istituto Nazionale di Fisica Nucleare, 
the French Centre National de la Recherche Scientifique, 
the French Commissariat \`{a} l'Energie Atomique, 
and the Korean Science and Engineering Foundation.  
The Southeastern Universities Research Association (SURA) operated the Thomas Jefferson
National Accelerator Facility for the United States Department of
Energy under contract DE-AC05-84ER40150.


\bibliography{THEBIB}

\end{document}